\documentclass[prl,aps,epsfig,twocolumn,showpacs]{revtex4}
\usepackage{graphicx}
\begin{document}

\title{ Continuously varying exponents in $A+B \rightarrow 0$ reaction with
long-ranged attractive interaction}
\author{Sungchul Kwon, S. Y. Yoon, and Yup Kim}
\affiliation{Department of Physics and Research Institute of Basic
Sciences, Kyung Hee University, Seoul 130-701, Korea}

\date{\today}

\begin{abstract}
We investigate the kinetics of the $A+B \rightarrow 0$ reaction
with long-range attractive interaction $V(r) \sim -r^{-2\sigma}$
between $A$ and $B$ or with  the drift velocity $v \sim
r^{-\sigma}$ in one dimension, where $r$ is the closest distance
between  $A$ and $B$. It is analytically show that the dynamical
exponents for density of particles ($\rho$) and the size of
domains ($\ell$) continuously vary with $\sigma$ when $\sigma <
\sigma_c =/1/2$, while that for the distance between adjacent
opposite species ($\ell_{AB}$) varies when $\sigma <
\sigma_c^{AB}= 7/6$. Beyond $\sigma_c^{AB}$, diffusive motions
dominate the kinetics, so that the dynamical behavior for
diffusive systems is completely recovered. These anomalous
behaviors with the two crossover values of $\sigma$ are supported
by numerical simulations and the argument of effective repulsion
between the opposite species domains.

\end{abstract}

\pacs{05.70.Ln,05.40.-a} \maketitle


The irreversible two-species reaction $A+B \rightarrow 0$ has been
intensively and widely investigated as a basic model for various
phenomena in physics \cite{first,phy}, chemistry \cite{chem}, and
biology \cite{bio}. When an $A$ and a $B$ particle meet on the
same site, they instantly and irreversibly combine to form an
inert species. Until now the studies for the reaction have been
focused on understanding the effect of the fluctuations of the
initial particle density or the global bias of particle motions on
the kinetics without careful consideration of the interactions
between $A$ and $B$ even for charged particles
\cite{first,monopole,kang,zumofen,namb,length,rg,hardcore}.
However, in situations where the kinetics of reaction is affected
by certain attractive interactions between $A$ and $B$ such as the
Coulomb interaction, the interactions should be much more
important than the simple diffusive motions or global biases to
understand the kinetics. Such situations may include
matter-antimatter annihilation in the universe,
soliton-antisoliton recombination, charge recombination in clouds
\cite{soliton} and electron-hole recombination in irradiated
semiconductor structures \cite{length}.

\begin{figure}
\includegraphics[width=5cm]{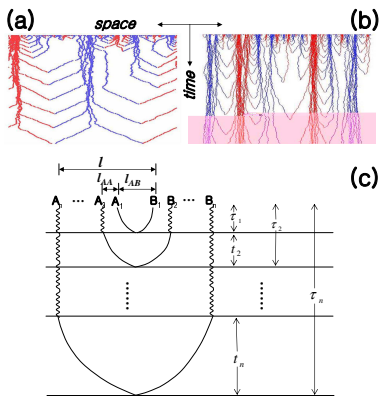}
\caption[0]{\label{pattern} (Color online) Snapshot of
trajectories of $A+B \rightarrow 0$ with the attractive
interaction between $A$ and $B$ of (a) $\sigma=0$ and (b)
$\sigma=0.3$. (c) The magnified schematic trajectories of adjacent
opposite species domains. Subscripts \{1,2,...,n\} indicate the
order of the positions of particles from a given domain boundary.
}
\end{figure}

It has recently been shown through a simple model that the
underlying attractive interaction between opposite species leads
to completely different scaling behaviors from those studied so
far \cite{ab-arw1}. In the model, the fluctuation-dominated
kinetics leads to the segregation of alternating $A-$rich and
$B-$rich domains as usual \cite{kang}. In addition,
domain-boundary particles feel the attractive interaction and are
assumed to move to the adjacent opposite species domain with a
constant drift velocity. Since the bias is a constant regardless
of the distance between $A$ and $B$, the interaction strength in
the model \cite{ab-arw1} is {\it infinite}. In contrast the
particles inside the domain are screened by the same neighboring
particles and the motion of bulk particles is naturally assumed to
be isotropic diffusion. As a result, the interaction causes the
alternatively changing bias at domain boundaries which is neither
the relative nor the uniform bias of Refs. \cite{kang} and
\cite{hardcore} (Fig.~\ref{pattern}(a)). In non-interacting
systems, the density decay has been known to depend on the motion
and the mutual statistics of particles. For isotropic diffusions,
the particle density $\rho(t)$ scales as $\rho (t) \sim t^{-d/4}$
in $d$ dimensions ($d \leq 4$)
\cite{monopole,kang,zumofen,namb,length,rg}. With the global
relative drift, $\rho (t)$ scales as $\rho (t) \sim t^{-(d+1)/4}$
for $d \leq 3$ \cite{kang}. With the uniform drift of both
species, the hard-core (HC) constraint between identical particles
leads to the scaling of $\rho(t) \sim t^{-1/3}$ in one dimension
\cite{hardcore}. With the infinite interaction, $\rho (t)$ scales
as $t^{-d/3}$ regardless of the HC constraint \cite{ab-arw1}. In
one dimension, the uniform bias of HC particles and the infinite
interaction lead to the same scaling law, but the scaling
behaviors of basic lengths are different. Hence the infinite
interaction results in new dynamical scaling behaviors
\cite{ab-arw1}.

The constant drift or the infinite attractive interaction is
unnatural \cite{ab-arw1} and cannot explain more general or real
situations where the attractive interaction depends on the
distance $r$ between $A$ and $B$. Some of physically realistic
attractive interactions should be those described by a conserved
attractive potential $V(r)\sim - r^{-2\sigma}$. The drift velocity
$v$ is then given as $v(r) \sim \sqrt{|V|} \sim r^{-\sigma}$. In
this paper, we investigate the kinetics of $A+B \rightarrow 0$
with the attractive potential $V(r)\sim - r^{-2\sigma}$ or with
the drift velocity to opposite species $v(r)=r^{-\sigma}$. In
reality the motions of boundary particles cannot be determined
only by the interaction, because there should exist various
noises, which are normally believed to make the background
diffusive motions. In this reason, a boundary particle is assumed
to stochastically move to its opposite species domain with the
rate $p=(1+v(r))/2$, and to its own domain with $1-p$ for a unit
time. Hence this model naturally includes the competition between
the attractive interaction and the diffusive motion. The drifted
motions at boundaries and the competition are expected to lead
rich and interesting scaling behavior as $\sigma$ varies.
Fig.~\ref{pattern} (b) shows the space-time trajectories for
$\sigma=0.3$. Using the trajectories schematically depicted in
Fig.~\ref{pattern} (c), we analytically derive asymptotic scaling
behavior. Intriguingly, we find that all dynamical exponents for
various densities and lengths continuously vary when $\sigma <
\sigma_c$. Furthermore the scaling behaviors for the present model
manifest unique anomalous behaviors, which have not seen in
dynamical critical models with the continuous varying exponents
\cite{conti}. In addition, the anomalous behavior is that there
exist two different crossover values of $\sigma_c$. $\sigma_c$ for
the density of particles ($\rho$) and the size of domains ($\ell$)
is $\sigma_c =1/2$, while that for the distance between adjacent
opposite species ($\ell_{AB}$) is $\sigma_c^{AB}= 7/6$. Hence the
interaction completely changes the scaling behavior of kinetics
when $\sigma < \sigma_c^{AB}$. However, when $\sigma \geq
\sigma_c^{AB}$, the kinetics for diffusive motions is completely
recovered \cite{length}. As in the $\sigma=0$ case, the HC
constraint is irrelevant due to the isotropic diffusions inside
domains. We also numerically confirm our analytical results.

With the equal initial density $\rho_A (0)=\rho_B (0)$, particles
are randomly distributed on an one-dimensional lattice of size
$L$. When the selected particle has two opposite species neighbors
such as ($A\cdots A \cdots B$), the central particle ($A$) hops to
the opposite species ($B$) with rate $p =(1+ v(r))/2$ and to the
same species ($A$) with rate $1-p$, where $v(r)=r^{-\sigma}$ with
$\sigma \geq 0$ and $r$ is the distance between $A$ and $B$.
Otherwise, particles diffuse isotropically. In the region of a
length $\ell$, the number of $A$ species is initially $N_A =
\rho_A (0) \ell \pm \sqrt{\rho_A (0)\ell}$ and the same for $N_B$.
After a time $t \sim \ell^z$, particles travel throughout the
whole of the region, and annihilate by pairs. The residual
particle number is the number fluctuation in the region, so we
have the relation $N_A \sim \sqrt{\ell}$ or $\rho_A \sim
1/\sqrt{\ell}$ for a given length $\ell$ \cite{monopole,kang}. As
the processes evolve, the system becomes a collection of
alternating $A$-rich and $B$-rich domains. To characterize the
structure of segregated domains, we introduce three length scales
as in Ref. \cite{length}. The length of the domain ($\ell$) is
defined as the distance between the first particles of adjacent
opposite species domains \cite{length}. The length $\ell_{AB}$ is
defined as the distance between two adjacent particles of opposite
species, while $\ell_{AA}$($\ell_{BB}$) is the distance between
adjacent $A$($B$) particles in a $A$($B$) domain. These lengths
scale asymptotically as
\begin{equation}
\label{exp-define}
\ell \sim t^{1/z},\;\; \ell_{AA} \sim
t^{1/z_{AA}},\;\; \ell_{AB} \sim t^{1/z_{AB}} \;\;.
\end{equation}

A bulk particle inside single species domains diffuses
isotropically and the mean position of the bulk particle is not
substantially changed until it becomes a boundary particle. Once
it becomes boundary, it drifts to its neighboring opposite species
with velocity $v(r)$ until it annihilates in the midway between
two opposite species domains. From this situation in mind we now
analytically calculate the scaling behaviors of the kinetics when
the drift velocity is $v(r) \sim r^{-\sigma}$. For an $AB$ pair
with the distance $\ell_{AB}$, the time interval for annihilation
of the $AB$ pair is $\tau_{AB} \simeq \ell_{AB} / v(\ell_{AB})
\sim \ell^{\sigma+1}_{AB}$. For $\sigma \geq 0$, the space-time
trajectory of a particle is arch-shaped. In the $\sigma=0$ case
the trajectory is very close to a pentagon \cite{ab-arw1}. These
arch-shaped trajectories should be self-similar (self-affine)
fractal structures, because they should have the scaling symmetry
due to the power-law scaling behavior (1). A typical base unit of
the self-similar arch-shaped trajectories of adjacent opposite
domains are schematically depicted in Fig.~\ref{pattern}(c). This
base unit allows us to calculate a time $\tau_\ell$ needed to
remove the unit of size $\ell$ surrounded by one scale larger
ones. Then the size of the larger unit increases by $\ell$ during
$\tau_\ell$ and we have
\begin{equation}
\label{ell}
 d \ell /dt \sim \ell /\tau_{\ell} \;,
\end{equation}
which gives the dynamic exponent $z$. In following calculations,
we consider only the mean positions of bulk particles, and assume
$\ell_{AA} (t)$ to be a constant during the annihilation of the
base unit. After a smaller unit is completely annihilated, the
remainder of particles redistribute over the larger unit increased
by the size of the annihilated unit. Hence we approximate
$\ell_{AA}(t) = \cdots = \ell_{AA}(t+\tau_n)= \cdots =
\ell_{AA}(t+\tau_\ell )$ during the annihilation of a smaller
unit.

As only boundary particles of each domain have two opposite
species neighbors, the annihilation of boundary particles comes
from the attractive interaction. It takes a time $\tau_1 =
\ell^{\sigma+1}_{AB}$ for the first boundary pair, $A_1$ and $B_1$
in Fig.~\ref{pattern} (c), to annihilate. The second pair ($A_2$
and $B_2$) from the boundary isotropically diffuse during time
$\tau_1$ until $A_1$ and $B_1$ annihilate. After the time
$\tau_1$, the second pair becomes a new boundary pair, and the
$t_2 \sim (\ell_{AB} + \ell_{AA})^{\sigma+1}$ is needed for the
annihilation. So it takes time $\tau_2 \sim \tau_1 + t_2$ in total
for the second pair to annihilate. Similarly, the $n$th pair from
the initial boundary will annihilate after $\tau_n \sim \tau_{n-1}
+ t_n$, where $t_n = (\ell_{AB} + (n-1) \ell_{AA})^{\sigma+1}$ for
$n \geq 2$. From the recurrence relation of $\tau_n$, we find
\begin{equation}
\begin{array}{lll}
\tau_n &\sim& \ell_{AB}^{\sigma+1} +
\sum^{n-1}_{k=1}(\ell_{AB}+k\ell_{AA})^{\sigma+1} ,\\
\\
&\sim& \ell^{\sigma+1}_{AB} +
[(\ell_{AB}+n\ell_{AA})^{\sigma+2}-\ell^{\sigma+2}_{AB}]/\ell_{AA}
\;.
 \end{array}
\end{equation}
The second line is obtained by integrating out the summation over
$k$. As the number of particles $N_{\ell}$ and the length
$\ell_{AA}$ in a domain of size $\ell$ scale as $N_\ell \sim
\sqrt{\ell}$ and $\ell_{AA} \sim 1/\rho \sim \sqrt{\ell}$, the
time $\tau_\ell$ needed to annihilate the domain of size $\ell$ in
the base unit is given by
\begin{equation}
\tau_\ell \sim \ell^{\sigma+1}_{AB} + [(\ell_{AB}+
\ell)^{\sigma+2}-\ell^{\sigma+2}_{AB}]/\sqrt{\ell} \; .
\end{equation}
Because of $\ell > \ell_{AB}$ by definition, we finally find the
leading scaling of $\tau_\ell$ as
\begin{equation}
\tau_\ell \sim \ell^{\sigma +3/2} \;\;,
\end{equation}
and $z$ is given as $z = \sigma +3/2$ from Eq.~(\ref{ell}).
Intriguingly, $z$ increases continuously and linearly with
$\sigma$. However $z$ cannot increase beyond the upper bound
$z_c=2$, because domains cannot spread more slowly than those of
random diffusion. Hence for $\sigma \geq \sigma_c = 1/2$, we have
$z=2$ regardless of $\sigma$. The scaling of $\ell_{AA}$ is easily
obtained from the relation $\ell_{AA} \sim \sqrt{\ell} \sim
t^{1/z_{AA}}$. One finds $z_{AA} =2z= 2 \sigma +3$. Since
$\ell_{AA}$ with the diffusion only scales as $\ell_{AA} \sim
t^{1/4}$ \cite{length}, the critical value of $\sigma$ is also
$\sigma^{AA}_c = 1/2$. For $\ell_{AB}$, we consider the change of
density during time $\tau_{AB}$ \cite{length}. During $\tau_{AB}$,
one pair of $AB$ particle annihilate only between boundaries, the
change of particle density is given as  $d \rho / dt \sim -
\rho_{AB}/\tau_{AB}$, where $\rho_{AB}$ is the density of $AB$
pairs.. Using relations, $\rho \sim 1/\sqrt{\ell}$, $\rho_{AB} =
1/\ell$, and $\tau_{AB} \sim \ell^{\sigma+1}_{AB}$, one finds
$\ell_{AB} \sim t^{1/z_{AB}}$ with $z_{AB} = z= \sigma+3/2$.
Interestingly, $\ell_{AB}$ follows the scaling of $\ell$. However,
$\ell_{AB}$ with the diffusion only scales as $\ell_{AB} \sim
t^{3/8}$ \cite{length} and thus $\sigma_c$ of $\ell_{AB}$ is
$\sigma^{AB}_c = 7/6$. The scaling exponents of three lengths and
their critical values of $\sigma$, beyond which the diffusive
scaling behaviors recover, are summarized as follows.
\begin{equation}
\label{z}
\begin{array}{lllll}
 z=\sigma+3/2 \;\;&,& z_{AA} = 2z \;\;&,& z_{AB}
= z \;\;, \\
\sigma_c = 1/2 \;\;&,& \sigma^{AA}_c = 1/2 \;\;&,& \sigma^{AB}_c =
7/6 \;\;.
\end{array}
\end{equation}

From the scaling of lengths, asymptotic decays of various
densities can be extracted. The density of total particles ($\rho
= 2\rho_A$), adjacent pairs of same species ($\rho_{AA} =
\rho_{BB}$) and adjacent pairs of opposite species ($\rho_{AB}$)
scale as
\begin{equation}
\rho \sim t^{-\alpha} ,\; \rho_{AA} \sim t^{-\alpha_{AA}} ,\;
\rho_{AB} \sim t^{-\alpha_{AB}} \;.
\end{equation}
As $\rho \sim 1/\sqrt{\ell}$, we have $\rho \sim t^{-1/2z}$ with
$\alpha = 1/2z = 1/(2\sigma+3)$. $\rho_{AA}$ follows the same
scaling of $\rho$ due to $\rho_{AA} \sim 1/\ell_{AA} \sim
1/\sqrt{\ell}$ so $\rho_{AA} \sim t^{-\alpha}$ with $\alpha_{AA} =
\alpha$. Finally $\rho_{AB}$ scales as $\rho_{AB} \sim 1/\ell$,
which leads to $\rho_{AB} \sim t^{-1/z}$ with $\alpha_{AB} = 1/z =
1/(\sigma+3/2)$. As densities with the diffusion only scale as
$\rho \sim \rho_{AA} \sim t^{-1/4}$ and $\rho_{AB} \sim t^{-1/2}$,
the upper bound of $\sigma$ is $\sigma_c = 1/2$ for all densities.
All decay exponents of various densities are simply given as
\begin{equation}
\label{rho} \alpha = 1/2z, \;\; \alpha_{AA}=\alpha, \;\;
\alpha_{AB} = 1/z \;\;,
\end{equation}
and the critical value of $\sigma$ is $\sigma_c = 1/2$ for all
densities. For $\sigma=0$, the all scaling behaviors of the
constant drift are fully recovered \cite{ab-arw1}. As shown in
Eqs. (6) and (7), there exists two different crossover values of
$\sigma$, $\sigma_c(=\sigma_c^{AA}=1/2$ and $\sigma_c^{AB}(=7/6)$,
where $\sigma_c < \sigma_c^{AB}$. It is very peculiar that there
exists two crossover values. However the inequality $\sigma_c <
\sigma_c^{AB}$ directly comes from the anomalous scaling of
$\ell_{AB}$ with diffusion only. With diffusion only, adjacent
opposite domains are effectively repulsive by preferential
annihilations of nearby $AB$ pairs. Hence, opposite species pairs
are further apart than the typical interparticle distance
$\ell_{AA}$ and, as a result, $\ell_{AB}$ increases anomalously as
$t^{3/8}$ \cite{length}. This anomalous scaling of $\ell_{AB}$
leads to another crossover in addition to the crossover by
diffusion.

As the strength of the interaction becomes weak, the effect of the
random fluctuation by diffusion becomes strong. The competition
between the fluctuation by diffusion and the drift by the
interaction leads to the continuously decaying exponents and the
critical values of $\sigma$ above which diffusive motions dominate
the kinetics. On the other hand, since the interaction maintains
the Galilean invariance of the domain structure, the predictions
(\ref{z}) and (\ref{rho}) are expected to be independent of the HC
constraint.

\begin{figure}
\includegraphics[width=6cm]{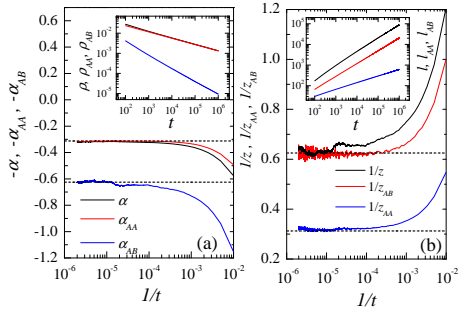}
\caption{\label{alpha01} (Color online) Effective exponents of
densities (a) and lengths (b) of the model with HC constraint for
$\sigma=0.1$. In each panel, two horizontal lines from top to
bottom show the predictions (a) $\alpha=0.313$ and
$\alpha_{AB}=0.63$ (b) $1/z=0.625$ and $1/z_{AA}=0.313$.}
\end{figure}

To confirm our analytic results numerically, we now present the
simulation results. With $\rho_A(0) = \rho_B(0)$, $A$ and $B$
particles distribute randomly on a chain of size $L$. In the
simulations we consider both HC particles and the particles
without the HC constraint, which we call bosonic particles. In the
model with HC particles there can be at most one particle of a
given species on a site. In the bosonic model there can be many
identical particles on a site. All the simulations are done on the
chain of size $L = 3 \times 10^6$ and with $\rho_A(0) = \rho_B(0)
= 0.1$. We average densities and lengths over $100$ independent
runs. Fig.~\ref{alpha01} (a) shows densities (inset) and their
effective exponents defined as $- \alpha (t)=
\log[\rho(2t)/\rho(t)]/\log 2$ for $\sigma = 0.1$ case of HC
particles. We estimate $\alpha = 0.317(5)$, $\alpha_{AA} =
0.3125(25)$, and $\alpha_{AB} = 0.625(25)$ respectively.  All
results agree well with the prediction (\ref{rho}); $\alpha =
\alpha_{AA} = 0.3125$ and $\alpha_{AB}=0.625$ for $\sigma=0.1$.
Fig.~\ref{alpha01} (b) shows the various lengths and their
effective exponents defined as$1/z(t) =
\log[\ell(2t)/\ell(t)]/\log 2 $ for $\sigma=0.1$. We estimate $z =
1.60(5)$, $z_{AA} = 3.20(5)$ and $z_{AB}=1.60(5)$ which also agree
very well with the prediction (\ref{z}); $z = 1.6$, $z_{AA} = 3.2$
and $z_{AB} = 1.6$ for $\sigma=0.1$. For bosonic particles, we
also obtain nearly the same results.

\begin{figure}
\includegraphics[width=6cm]{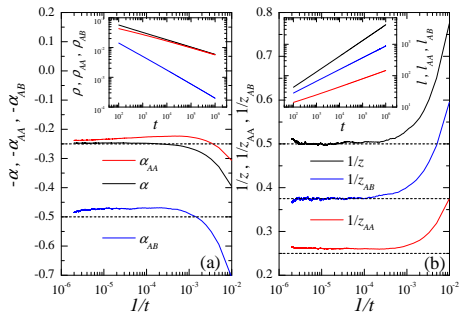}
\caption{\label{alpha125} (Color online) Effective exponents of
densities (a) and lengths (b) of the model with HC constraint for
$\sigma=1.25$. In each panel, horizontal lines from top to bottom
show the prediction (a) $\alpha = \alpha_{AA}=1/4$ and
$\alpha_{AB}=1/2$ (b) $1/z=1/2$, $1/z_{AB} = 3/8$ and
$1/z_{AA}=1/4$.}
\end{figure}

To check the crossover from the continuously varying dynamic
scaling to the completely diffusive one, we perform Monte Carlo
simulations under the same initial conditions for $\sigma = 1.25
(> \sigma_c^{AB} > \sigma_c)$. Fig.~\ref{alpha125} shows
simulation results of HC particles for $\sigma = 1.25$. We
estimate $\alpha = 0.25(1)$, $\alpha_{AA} = 0.24(1)$ and
$\alpha_{AB} = 0.47(1)$, which agree reasonably well with the
values without the drift in one dimension;
$\alpha=\alpha_{AA}=1/4$ and $\alpha_{AB} =1/2$ \cite{length}. For
lengths, we estimate $1/z=0.50(5)$, $1/z_{AA} = 0.26(1)$, and
$1/z_{AB}=0.375(5)$ which also agree well with the values without
the drift, $1/z=1/2$, $1/z_{AA} = 1/4$ and $1/z_{AB} = 3/8$
\cite{length}.

\begin{figure}
\includegraphics[width=6cm]{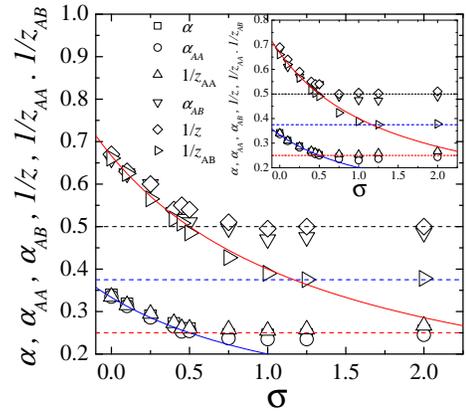}
\caption{\label{all}(Color online) Plot of dynamical exponents
versus $\sigma$. Main (inset) plot show the exponents of HC
(bosonic) particles. Solid lines in each panel from top to bottom
correspond to the predictions; $1/z_{AB}=1/(\sigma+3/2)$ and
$1/z_{AA}=1/(2\sigma+3)$. Horizontal dashed lines from top to
bottom correspond to $1/z=1/2$, $1/z_{AB}=3/8$, and $1/z_{AA}=1/4$
respectively.}
\end{figure}

We plot the estimates of all exponents versus $\sigma$ and the
lines for the predictions (\ref{z}) and (\ref{rho}) both for HC
particles and for bosonic particles in Fig. 4. All exponents
except $1/z_{AB}$ continuously vary along the predicted lines for
$\sigma < \sigma_c=1/2$. In contrast $1/z_{AB}$ varies along the
predicted line for $\sigma < \sigma_c^{AB}=7/6$. Beyond $\sigma_c$
(or $\sigma_c^{AB}$), each exponent takes the value of the
diffusive system without the interaction regardless of $\sigma$.
The exponents for HC particles are nearly identical to those for
the bosonic particles, and this result comes from the irrelevance
of the HC constraint to the kinetics due to the isotropic
diffusion inside domains.

In conclusion, we investigate the kinetics of irreversible
reaction $A+B \rightarrow 0$ with the attractive interaction $V(r)
\sim -r^{-2\sigma}$ between $A$ and $B$ or with the drift velocity
$v(r)\sim r^{-\sigma}$ to opposite species at segregated domain
boundaries. The drift leads to arch-shaped space-time trajectories
of particles with which we analytically derive asymptotic scaling
behaviors, and numerically confirm them. Intriguingly, the drift
results in continuously varying scaling behavior by certain
critical values, $\sigma_c$ or $\sigma_c^{AB}$. above which
diffusive motions dominate the kinetics. In our derivations, we
only consider the mean positions of particles and neglect the
expansion of domains by the random fluctuations during the
annihilations of domains. Hence when the interaction dominate the
kinetics, the fluctuation does not affect the asymptotic scaling
behavior.

This work is supported by Grant No. R01-2004-000-10148-0 from the
Basic Research Program of KOSEF.

\end{document}